\documentclass[intlimits,twoside,a4paper]{article}

\usepackage[T2A]{fontenc} 
\usepackage[utf8]{inputenc} 
\usepackage[greek,ukrainian,english]{babel}
\usepackage{cmpj3}

\usepackage{amsmath}
\usepackage{empheq}
\usepackage{pstricks}
\usepackage{mathrsfs}
\usepackage{mathtools}
\usepackage{cancel}
\usepackage{accents}
\usepackage{dsfont}
\usepackage{color}
\usepackage[normalem]{ulem}
\usepackage[caption = false]{subfig}

\usepackage{upgreek}

 \DeclareRobustCommand{\augiefamily}{%
 	\fontfamily{augie}\fontseries{m}\fontshape{n}\selectfont}
 \DeclareTextFontCommand{\textaugie}{\augiefamily}

\def\be{\begin{equation}}
	\def\ee{\end{equation}}
\def\bea{\begin{eqnarray}}
	\def\eea{\end{eqnarray}}
\def\bpar{\left(\!\!\begin{array}}
	\def\epar{\end{array}\!\!\right)}
\def\bdar{\left|\!\!\begin{array}}
	\def\edar{\end{array}\!\!\right|}
\def\barr{\begin{array}}
	\def\earr{\end{array}}
\def\btab{\begin{tabular}}
	\def\etab{\end{tabular}}

\def\<{\langle}
\def\>{\rangle}

\issue{2024}{27}{3}{33804}
\doinumber{10.5488/CMP.27.33804}

\title[A starter toolkit for researchers to explore early Irish literature]{Crossing the disciplines --- a starter toolkit for researchers who wish to explore early Irish literature}

\author[M. McCarthy, D. P. Curley]{M. McCarthy\orcid{0009-0005-2252-554X}, D. P. Curley\orcid{0000-0001-9495-4902}}
\address{
		Rathcroghan Visitor Centre, Ireland
		}

\date{Received March 15, 2024, in final form May 27, 2024}

\begin{document}
\maketitle
\sloppy
\begin{abstract}
				
The inspiration behind this paper came from both authors' long-term collaboration with our friend and colleague, Professor Ralph Kenna. This connection emerged initially through his interest in Rathcroghan and in our paper, `Exploring the Nature of the Fr\'aoch Saga', which we concluded with the statement that we believed it `presents a case that will hopefully ignite conversation between disciplines'.
This led us to consider the potential value for researchers of compiling a template list of useful and reliable sources and resources to consult, in other words a type of starter toolkit or guide for any individual from an alternative discipline or background, who might possess, or, in time, develop a personal or professional interest in Early Ireland and Early Irish literature.
In doing this, we decided for ease of illustration, to take the example of the location name Rathcroghan/Cruachan A\'i, (the prehistoric Royal Site of Connacht in the west of Ireland and the place that we both work in and interact with on a daily basis), as a case study in order to demonstrate an initial methodological approach to not only the types of resources and information available, but also to highlight some
potential pitfalls that may arise in the course of an investigation.

		 
\end{abstract}



\section{Introduction}	\label{I}

The inspiration for this short essay came from both authors' long-term collaboration with our friend and colleague, Professor Ralph Kenna. This connection emerged initially through his interest into our paper, `Exploring the Nature of the Fr\'aoch Saga', published in 2018, in which we considered the apparent connections between the mythological Ulster Cycle figure Fr\'aoch mac Fidach Foltruad, and a number of prominent archaeological and topological features in Mag nA\'i or Machaire Connacht (the Plains of Connacht, mid-Roscommon, Ireland)~\cite{McCarthy2018}. 

We had concluded this paper with the statement that we believed it \emph{`presents a case that will hopefully ignite conversation between disciplines'} and continuing on from this we had a very fruitful collaboration with Ralph and his colleagues on numerous studies and initiatives, including an undertaking with colleagues from Coventry University and the University of Limerick~\cite{Janickyj2022}, and not least our contributions to Ralph's pioneering \'EIR\'I (Evoking Ireland’s Resilient Female Icons) project.

Throughout his career, Ralph bridged the apparent divide between science and humanities, and the insights and scientific analysis that he brought to bear on our own investigations led to hitherto unconsidered research tools for us, which has undoubtedly lent an extra dimension to our work, one that we consider to be of enormous benefit and insight. This has made us consider, operating in the humanities side of the discourse, the potential value for researchers of compiling a template list of useful and reliable sources and resources to consult, in other words a type of starter toolkit or guide for any individual from an alternative discipline or background, who might possess, or, in time, develop a personal or professional interest in Early Ireland and Early Irish literature.

Therefore, in echoing Ralph's ethos, in a world of ever more specialised disciplines and areas of expertise where the need for collaboration and partnership is perhaps all the more necessary, crossing those apparent modern day divisions of the sciences and humanities is vitally important. 

Many great scholars of the past observed no real distinction between these subjects and, despite opposition, rightfully embraced all disciplines as worthy of consideration. One example of this is the 9th century philosopher John Scottus Eriugena (John the Irishman), who spent time in France at the invitation of King Charles the Bald, and whose seminal work, \emph{Periphyseon: On the Division of Nature} was placed on an Index of Forbidden Books by the Church~\cite{Rathcroghan2008}. It is therefore a particular tribute to Ralph that he, in a way, continued in the tradition of his fellow countryman to the benefit of all concerned. His work with his colleagues in network theory tackled not only aspects of early Irish literature, history and tradition, with its first application in medieval Irish literature when Ralph and Padraig MacCarron applied it to compare networks of interacting characters in Ireland's national epic tale, \emph{T\'ain B\'o Cuailnge}, the Anglo-Saxon \emph{Beowulf} and Homer's \emph{Iliad}~\cite{MacCarron2013}, but also the contemporary in, for example, investigation of the narrative structure in the modern epic, \emph{A Song of Ice and Fire}~\cite{Gessey2020}. In looking back to the past and bringing a fresh perspective to bear on early texts, stories and ideas, Ralph was, in reality, honouring a great scientific and human tradition and so, in kind, to honour him, we offer this paper as an initial resource and guide for those who may be inspired to follow in his footsteps.

In doing this, we decided for ease of illustration, to take the example of the location name Rathcroghan/CruachanA\'i\footnote{Rathcroghan --- \emph{Cruachan A\'i} --- the prehistoric capital of Connacht, Ireland’s western province, and the landscape upon which we work daily and which Prof. Kenna held a great interest in.} as a case study in order to demonstrate an initial methodological approach to not only the types of resources and information available, but also to highlight some potential pitfalls that may arise in the course of an investigation. To achieve our aim, we will give a brief outline and overview of the Rathcroghan archaeological landscape, early Irish medieval texts, some potential pitfalls that may be encountered and some common popular resources that are readily available to researchers.

\section{What is Rathcroghan/Cruachan A\'i?} \label{II}

When you begin the process of identifying even a seemingly singular entity such as Rathcroghan, the researcher must still practice caution when pursuing attestations in literary or historical sources. In the manuscript tradition, the modern anglicised name of Rathcroghan is actually traditionally referred to by any of a number of different names, which all refer to the same place. These include: \emph{Cruachu, Cruachan A\'i, Mag Cruachan, R\'ath Cruachain, S\'id Cruachain} amongst others~\cite{MacKillop1988}. These names, as with many places in Ireland, can have a topographical or geographical explanation, or conversely a more fantastical, mythological root, or as we shall see in the case of Rathcroghan, sometimes both together.

{\bf Cruach\'an} can be translated to numerous meanings --- a stack, pile, peak, hill, mound, tumulus, a conical mountain (such as \emph{Cruach P\'adraig/Cruachan Aigle} --- Croagh Patrick, Co. Mayo). However, alongside the toponymic features referenced above, in literary mythological references, two medieval texts, the \emph{C\'oir Anmann}\footnote{Fitness of Names.}  and the \emph{Rennes Dindshenchas} \footnote{Dindshenchas --- Lore of Place Names.} refer to Cruachan being derived from the figure of \emph{Cruacha Cr\'o Dhearg} (Crochen the Red-Skinned), who was said to be the mother of Queen Medb, the legendary queen of Connacht, and one of the main protagonists in \emph{An Táin Bó Cuailnge}~\cite{Curley2023}.

{\bf A\'i} --- The second part of the name, `A\'i', can be interpreted to mean a swan, a herd, a region, tract, country, territory, patrimony or inheritance, a cause, a dispute, wise or learned. However, in a similar fashion to Cruachan above, \emph{A\'i} may have a more fantastical derivation in referring to the liver of the great white
horned bull \emph{Finnbhennach}, an organ which was said to be shed at this location by the \emph{Donn Cuailnge} (the Brown Bull of Cooley) at the close of \emph{T\'ain B\'o Cuailnge}. These two magical bovines were the prizes fought over in the latter epic~\cite{Curley2023}.

Combining the two terms in translation could yield something akin to \emph{Cruachan A\'i} --- the mound of the plain of A\'i.

{\bf Rathcroghan} --- In considering the label of Rathcroghan; \emph{R\'ath Cruachan} literally translates as `The Fort of Cruachan' and is a name which is significant on many levels. In terms of archaeological remains, Rathcroghan, located to the north-west of the village of Tulsk in Co. Roscommon, Ireland, is a collection of 240 identified archaeological sites, contained within an area of 6.5~km$^2$, and which range in date from the Neolithic Period to the late Medieval Period, spanning over 5,500 years of history. It is the location of some 37 identified burial mounds from the Bronze and Iron Age, numerous ringforts (settlement sites), standing stones, linear earthworks, stone forts, a great Iron Age ritual sanctuary, and a cave that is described in Irish medieval literature as an Otherworldly entrance and even on occasion as a Gate to Hell~\cite{Waddell2014}.

As might be expected then, Rathcroghan is also incredibly important from a literary point of view and it is recorded as one of the locations for large-scale ceremonial assembly or \emph{\'oenach} in Ireland~\cite{Waddell2009}. These assemblies took place at important transition points in the year, the changing of the seasons, and were occasions for judgements to be passed, for kings to be inaugurated and great feasting and entertainments. Rathcroghan is also recorded as one of the three chief burial places of Ireland, alongside the Fair of Tailtiu and Br\'u na B\'oinne~\cite{Waddell2009}. Moreover, in several early tales, Rathcroghan figures as a kingly settlement for the Connachta (descendants of Conn), the ruling dynasty in the territory of Connacht, the western province of Ireland, from about the fifth century. 

As alluded to above, Rathcroghan features very heavily in the mythological Ulster Cycle of Tales (see below), particularly as it is the location of the palace of the Warrior Queen Medb (Maeve) of Connacht, and therefore the central tale of this cycle, Ireland’s national epic, \emph{An T\'ain B\'o C\'uailnge} (The Cattle Raid of Cooley), locates both its beginning and end in the Rathcroghan landscape. Moving into the late Medieval Period, Rathcroghan still retained a symbolic hold over the elites of Ireland, and much evidence exists to show that it continued to be regarded as synonymous with the kingship of Connacht.
An interesting point to note within this is the context in which a reference to a place like Rathcroghan is taken in, as is illustrated in the medieval text of \emph{F\'eilire Aonghusa} or Martyrology of Aonghus\footnote{Referring to Aonghus of Tallaght (a prominent C\'eil\'i D\'e/Client of God)}~\cite{ORiain2011}, wherein the author appears to take delight and satisfaction at the demise of Rathcroghan in lieu of the contemporary thriving of Christian centres. An important question to pose in this context, rather than simply taking the author at face value, is to consider that there could be a double meaning contained within this, in that an earlier important, pre-Christian site held a level of societal significance so as to be still  worthy of the writer’s attention, centuries after Christianity arrived on the island.

\begin{quotation}
	\centering{\bf \emph{`The great settlement of Tara has died with the loss of its princes; \\
	Great Armagh lives on with its choirs of scholars... \\
	The fortress of Cr\'uachain has vanished with Ailill, victory’s child; \\
	A fair dignity greater than kingdoms is in the city of Clonmacnoise...'}}~\cite{Waddell2009}
\end{quotation}

\section{Early Irish literature} \label{III}

The coming of Christianity and the Latin alphabet to Ireland in the fifth century coincided with the emergence of the first known written version of the Irish language in the form of og(h)am script, a writing technique consisting of strokes and notches along a stemline. Examples of ogham largely survive to us today through inscriptions in stone, and the system was in use between the 4th to the 6th/7th centuries AD~\cite{Stifter2006}.

Due to the lack of surviving evidence, prior to this, it is presumed that knowledge was passed down primarily in oral form, which in early Ireland was the domain of an extremely important section of Irish society known as the \emph{filidh} or poetic class. This tradition appears to have continued in parallel with the development through the monasteries of a system of written Irish, and whatever the conjectures on the interactions between the learned classes of Christian scribes and Gaelic poets in early Irish society, it seems that broadly between the seventh and twelfth centuries, a form of communion must have occurred between the two which produced a vast body of vernacular literature~\cite{MacCana1980}. This proposed process is perhaps highlighted in the 6th century example of the poet Colmán mac L\'en\'eni, who after becoming a monk, founded the monastery of Cloyne, but continued in his practice of poetry~\cite{Carney2005} and the collaboration is further superbly illustrated by the twelfth century text of \emph{Lebor G\'abala} or Book of Invasions, which blends the mythology and lore of Ireland with the biblical history of the world, beginning with Noah and the Great Flood. To give a sense of the extent of the early Irish narrative tradition, a cursory look at the catalogue of Ulster Cycle tales on the CODECS Van Hamel catalogue (see below) yields in excess of 190 individual surviving tales, while allowing for some slight duplication. Couple this with the fact that the application of the unseen species model of data analysis in respect of Irish provenanced-manuscripts determines that possibly as much as 80\% of medieval Irish literature still survives today for inspection~\cite{Kestemont2022}, meaning that there is a very extensive corpus of material available to the prospective researcher. 

For convenience sake this extensive collection of early Irish literature is often broken up into  four major `cycles'~\cite{Rolleston1994}, though it must be noted that these classifications were not employed by the medieval authors themselves, who tended to demarcate tales according to subject matter~\cite{Williams2016}. In adhering for our purpose here to the system of cycles, although there is overlap in places, they are usually referred to in turn as:
\begin{enumerate}
	\item {\bf The Mythological Cycle} --- this concerns itself with the coming of various deities, peoples and tribes to the island of Ireland, as recorded in the \emph{Lebor G\'abala} or Book of Invasions mentioned above, and which culminates in the arrival of the Milesians, who are purported to be the ancestors of the Irish~\cite{MacKillop1988}.
	\item {\bf The Ulster Cycle (An R\'ura\'iocht)} --- set in the Irish Iron Age, roughly around the era of Christ, this cycle concerns itself with the wars between the western province of Connacht and the northern province of Ulster and the enmity between their two rulers, Queen Medb of Connacht, who ruled from Rathcroghan and King Conchobar Mac Nessa of Ulster who resides with his Red Branch Knights in Emain Macha (in modern day county Armagh). Of the four cycles, from a literary standpoint this is undoubtedly the pinnacle of early Irish literature, and this group of verse and prose tales, transcribed as early as the eighth century, contains Ireland’s national epic --- \emph{An T\'ain B\'o Cuailnge} (the Cattle Raid of Cooley).
	\item {\bf The Fenian Cycle (An Fhianna\'iocht)} --- this concerns itself with the exploits of the legendary warrior band referred to as the Fianna and their leader, Fionn mac Cumhaill and they are set during the reign of the legendary king, Cormac mac Art in the 3rd Century A.D.~\cite{Rolleston1994}. The important 12th century text of \emph{Acallam na Sen\'orach} (The Tales of the Elders of Ireland) records two fenian warriors, Fionn mac Cumhaill’s son, \'Ois\'in and Ca\'ilte relating the exploits of the Fianna to Saint Patrick~\cite{MacKillop1988}.
	\item {\bf The Cycle of Kings or Historical Cycle} --- sometimes referred to as the historical cycle, this concerns itself with both mythical and historical kings and kingship from the 3rd to the 7th century A.D.~\cite{MacKillop1988} and contains such tales as the infamous \emph{Buile Shuibhne} (The Frenzy of Sweeney).
\end{enumerate}

\section{Some potential pitfalls} \label{4} 

In approaching early Ireland and its literature, there are, like in any other discipline, numerous potential pitfalls to be aware of, not least the fact that so-called `Celtic Culture' has seen a massive revival throughout the 19th, 20th and even 21st centuries. This trend is generally referred to under the heading of `The Celtic Twilight' and in Ireland it incorporated the cultural resurgence of both the mythology of Early Ireland and the Irish language in this period. Huge popularity was brought to these tales and characters by the poetry of figures like Sir Samuel Ferguson and W.B. Yeats and the writings of Lady Augusta Gregory. While this undoubtedly has been a significant movement in the preservation of Irish culture and language, it is essential to note that many examples of these writings are of their time, occasionally displaying a Victorian type constraint, and may also sometimes contain a certain romantic flavour of a yearning for an idealised, lost, idyllic, noble society, with many of the interpretations having an artistic rather than a scholarly influence. This is, of course, perfectly acceptable, but an important piece of information to note from the standpoint of the researcher, as many modern day compendiums and interpretations are still influenced by these writings and therefore may not always be suitable as a source for primary early material.

A notable instance of this type of influence and which has a relevance to our own case example of Rathcroghan, is the popular idea, which still holds true to this day, that the mythical figure of Queen Medb mentioned above, is said to be buried in the Neolithic burial cairn on top of the mountain of Knocknarea. This mountain on the C\'uil Irra peninsula in Co. Sligo is topped by this spectacular cairn of Measc\'an Mh\'eabha, measuring $c$.60~m in diameter and with an impressive height of $c.$10~m and is the dominant feature of a major Neolithic ritual landscape for the region~\cite{Bergh2000}. 
The prevalent notion for Knocknarea as Medb’s final resting place seems to have become particularly popularised during the Celtic Twilight by some of the figures mentioned above, such as Ferguson and Yeats, who in his poem the `Wanderings of Oisin', refers to Medb’s grave:

\begin{quotation}
	\centering{\bf \emph{`The cairn-heaped grassy hill, where passionate Maeve is stony still'}}
\end{quotation}

Aside from the fact that this is a Neolithic cairn and that the tales that surround Queen Medb are set in the Iron Age, around the time of Christ,\footnote{To quote Mallory \emph{`there is no simple date for the T\'ain'}~\cite{Mallory1992}. The literary composition suggests an Iron Age setting, with the death of Conchobar mac Nessa in the Ulster Cycle recorded in the 12th century Book of Leinster being said to coincide with the crucifixion of Christ on Calvary~\cite{Meyer2011}. Furthermore, the medieval annals record Medb’s father, Eochaidh Feidlach as \emph{Ard R\'i} (high-king) in the year 137 BC (Anno Mundi 5058)~\cite{CELT2}, though the language and material culture represented in the epic indicates that it has been remoulded to appeal to a later medieval audience, perhaps reflecting the contemporary authors literary attempt to portray a world of the past~\cite{Mallory1992}.}  we have also noted that Rathcroghan --- \emph{Cruach\'an A\'i} --- was once a great cemetery and so it is perhaps not surprising that the much earlier 12th century manuscript, \emph{Lebor na hUidre} (Book of the Dun Cow), which also contains the earliest written version of \emph{T\'ain B\'o Cuailnge}, contains a tract entitled \emph{Senchas na Relec} or the Lore of Burial Places, which records the inhumation of many luminaries in the Royal Cemetery of Cruach\'an, including most importantly, Queen Medb herself~\cite{Donovan1844}. Given the fact that the literary, mythological character of Medb is now generally viewed as an echo of an earlier territorial divinity and not an historical figure~\cite{Waddell2014} which implies that she may never have been buried at all, this example is nevertheless illustrative of how a popular contemporary notion can take hold and gain traction and credibility over time~\cite{McCarthy2024}. In considering this, it is also vital to note that when we view the writings of medieval church scribes that the same factors are at play, and that when we are considering these texts, we are not getting a pure insight into the past, but rather, as in the example of \emph{F\'eilire Aonghusa} noted above, the reflections and interpretations of a medieval mindset. 

These concepts may be given extra credence by, for example, the types of weaponry described in the \emph{T\'ain B\'o Cuailnge}, where items such as the swords described as being wielded by the warriors seem to resemble later contemporary medieval designs rather than what is evident from the Iron Age archaeology~\cite{Mallory2016}. Furthermore, the emphasis on the province of Ulster in the narrative is interesting, and could be explained with the theory that the version of the story as we have it today was first collated in this territory, which is consistent with compelling evidence that the saga was re-worked in what becomes Co. Louth in the 9th century, with particular emphasis given to local places and concerns~\cite{OhUiginn1992}. Therefore, while there may have been a time when these tales were popularly conceived as close to historical, a concept which can still persist in certain quarters, from the point of view of approaching it as a research topic, these are extremely important factors to be aware of.

\section{Toolkit} \label{5} 

So far in our case study we have provided an overview of Rathcroghan itself and referenced the important central role that it plays in early Irish epic literature, while also highlighting some potential pitfalls to avoid in any initial investigation. To continue this theme, it is useful to place a focus on some particular areas which would be considered useful to investigate, and alongside this to highlight some readily accessible and reliable resources that are available. In our case study of Rathcroghan, for example, given the broad scope of its historical and archaeological significance, it could be considered under some of the following source headings; literary texts, historical sources and annalistic materials, \emph{dindshenchas} (place name lore), praise poetry, and even toponymy.

From a literary standpoint, we have already highlighted the cycles into which early Irish literature is broadly divided, and the sources for these materials are the trove of early Irish manuscripts, the majority of which are held in numerous disparate institutions such as the Royal Irish Academy, Trinity College Dublin and various other academic bodies and private collections throughout Ireland, Britain and Europe. While accessing these vulnerable manuscripts directly is obviously not a practical endeavour, for those who wish to view them virtually, ISOS (Irish Script on Screen), provided by the Dublin Institute of Advanced Studies, holds copious digitised manuscripts and collections, which can be viewed by the public online~\cite{DIAS}.
In following this thread, for transcribed and translated texts, CELT (Corpus of Electronic Texts) is a free online digital resource for Irish literature and contains items such as annalistic and \emph{dindshenchas} material (see below) as well as literary and liturgical resources~\cite{CELT}. In a similar vein CODECS (Collaborative Online Database and e-Resources for Celtic Studies) is an ongoing project which is compiling a database of sources of interest to Celtic studies and includes texts, manuscripts and bibliographies~\cite{CODECS}. These two databases combined provide an essential and invaluable online resource as a starting point in elucidating texts and references on any item relating to early Ireland.
In taking our example of Rathcroghan, a simple search in the CELT database returns references to sources for the `Death Tales of the Ulster Heroes' from the 12th century Book of Leinster, a transcribed translation of the Rath Cruachan poem from the \emph{Metrical Dindshenchas}, a reference to the tribes of Connacht from the 15th century Glenmasan manuscript and extracts from Geoffrey Keating’s 17th century History of Ireland (\emph{Foras Feasa ar \'Eirinn}) to highlight but a few. In a similar fashion, an entry on CODECS will return an impressive data set of references and also available primary and secondary sources for each piece of material selected, with the relevant links to sites such as the aforementioned CELT and \url{https://archive.org/}, where these sources are available online.

As alluded to above, \emph{Dindshenchas} or place name lore is a collection of stories and legends in both metre and prose, relating to the toponymy of various places in Ireland, explaining, often in a fantastical fashion, the origins and meaning of various place names, with many of these explanations drawn from the literary cycles already highlighted. There is no complete surviving collection, but the main text is contained in the 12th century Book of Leinster~\cite{OhOgain2006} and \emph{dindshenchas} material is also located in numerous other Irish manuscripts alongside other collections, such as the \emph{Rennes Dindshenchas} in France~\cite{MacKillop1988}. Taking Rathcroghan in this instance, as mentioned above, the Rennes Dindshenchas explains Cruachan, not from a geographical standpoint, but rather as being derived from the mythical figure of \emph{Cruacha Cr\'o Dhearg} (Crochen the Red-skinned), who was purported to have been the mother of Queen Medb~\cite{Curley2023}

\begin{quotation}
	\centering{\bf \emph{`And hence is Raith Maighe Cr\'uachan, From Cr\'uachu, Et\'ain’s handmaid, (so called) Because her head was blood red, Together with her eyebrows and eyelashes'}}\\
	\hspace{27em}{Rennes Dindshenchas}
\end{quotation}

Another medieval source text, the C\'oir Anmann, agrees with this derivation~\cite{Curley2023}:

\begin{quotation}
	\centering{\bf \emph{`Medb of Cr\'uachu and Cr\'uachu itself, whence are they? Easy to say, Cr\'uachu that is Crochen the Red-skinned the handmaid of Medb’s mother \'Et\'ain, It is from her that Medb of Cr\'uachu and Cr\'uachu itself got their names'}}\\
	\hspace{29em}{C\'oir Anmann}
\end{quotation}

Alongside this \emph{dindshenchas} material, the Irish annals or yearly accounts, which were compiled in the scriptoria of medieval Irish monasteries, record events such as famous deaths of both nobility and laity, accounts of significant battles, plagues and miraculous cosmic events, and document a chronicle of Irish history, with some notable examples among the collections being: The Annals of Ulster, The Annals of Tigernach, The Annals of Connacht and the Annals of Clonmacnoise. Following on from these texts, produced in the 17th century under the guidance of M\'icha\'el \'O Cl\'eirigh, the Annals of The Four Masters opens in the year AM 2242\footnote{AM --- Anno Mundi --- the year of the world}, which they record as the year of the Flood, with continuous coverage following after the year AM 3266 and finally concludes in 1616 AD~\cite{Cunningham2014}, with entries becoming more reliable and realistic as they move further forward into the historical period.
In briefly considering Rathcroghan in the context of the Annals, the Four Masters records that in the year AM 3519: \emph{`At the end of these three years Muimhne died at Cruachain. Luighne and Laighne fell in the battle of Ard Ladhron by the sons of Emhear'}~\cite{CELT2}. The Annals of Tigernach records \emph{`Ailill son of Maga (who reigned at Cruachan with Medb) in 30 BC...'}~\cite{Mallory1998}, while moving into the later historical period, the Annals of Ulster record the promulgation of Saint Patrick’s and Saints Ciaran’s law at Cruachain, in the years 782 and 813 AD respectively~\cite{Mallory1998}.

Praise poetry is a different class of historical source to those discussed above, in that it was usually addressed to the lay nobility, was designed to be publicly recited, and dealt primarily with the patron’s present political ambitions, as opposed to anything ancient in character. However, in so doing, the composers of these verses borrow from tropes of ancient pedigree for their patrons, in an effort to create or maintain legacy ties, and demonstrate legitimacy for their lord. Again, owing to the reputation of Rathcroghan as a symbolic hub through the millenia, it frequently features in the stanzas of medieval bardic verse, indicating that Rathcroghan remained known and continued in its significance in the consciousness of these poets, patrons and indeed it could be argued, by society more generally, well into the later medieval period and early modern period. For one example of many, we can point to the late-sixteenth or early seventeenth century poem composed by the poet Gofraidh mac Briain Mac an Bhaird for Diarmaid \'O Conchubhair Donn, beginning \emph{Congaibh riot, a R\'aith C[h]ruachan} (Hold fast, O R\'aith Chruachan...)~\cite{Hoyne2011}.

Finally, one of the best preserved, yet underused, sources that could be consulted for these endeavours is the toponymical or placenames record. The survival of townland names and the names of local features is an invaluable key to the former organisation of the early historic landscape, and the societies that utilised them~\cite{MacShamhrain1991,Bhreathnach2014}. A considerable amount of information can be gleaned from analysis of the toponymy, particularly townland names, as they are one of the primary forms of recording and remembering landscape~\cite{Kilfeather2010}, a use that is often overlooked. Although evidence for the origins of the townland system is difficult to confirm, it is in place by at least the twelfth century, as townlands are referred to in documentary sources of that date~\cite{OhAisibeil2018}. The divisions and in many cases names, can be presumed to be of considerable antiquity, with possible origins for the townland system in the early medieval period or even earlier~\cite{McErlean1983,Nicholls2003}. The Placenames Database of Ireland, accessed via \url{www.logainm.ie}, is the primary resource for searching recorded and translated Irish place-names. 

From the perspective of the potential value of place name-related enquiries, the investigations into the former significance of an aforementioned heroic figure from the Ulster Cycle of Tales is relevant. Fr\'aoch mac Fidach Foltruad, a notable warrior in Medb’s expeditionary army to Cooley in search of the \emph{Donn Cuailnge}, receives a relatively light treatment in the literary sources. However, conversely, his name is to be found at numerous locations across the mid-Roscommon area, associated with prominent archaeological monuments and landscape features. It is this disparity between literary survival and toponymic survival which first drew the attention of the present writers, and the reality that while the literary sources are invaluable, they are limited by the fact that they were created and thereafter accessible to only a slender segment of the medieval Irish population. The name survivals in the landscape are likely to present with evidence of the relevance of, in this case, a heroic figure, amongst society at large, and in certain instances, these relics of former times could be the critical first key into a fascinating journey of discovery, and which, in our case led to the friendship and collaboration across disciplines which occurred between Ralph and his colleagues, and ourselves in the humanities.

\section{Conclusion}

In conclusion, in our overview, we have seen the importance of Rathcroghan/Cruachan A\'i not only in archaeological terms, but also its historical and literary aspects, and as we proposed throughout the course of this essay, its prominence in early Irish literature is also very likely a reflection of its prominence in the medieval mindset of the scribes who are describing it.
We also pointed out some potential pitfalls and common misconceptions in approaching a subject of this type, as in the modern day information age, across all disciplines and subjects of study and research, there exists a corresponding body of misinformation which it is vital to be aware of, and Early Ireland is no exception in this.
In the production of this paper, it occurs to the authors, that in looking at the case study example of Rathcroghan above and in both elucidating information and proposing the separation of potential facts from fiction, that the application of scientific techniques and methodology, such as those Prof. Kenna brought to bear throughout his career, could bring a further standardisation of procedure for investigation that could only prove beneficial to all concerned and that the crossing of modern day boundaries between disciplines in all directions, could have the potential to bring about fresh viewpoints and approaches, perhaps not previously considered.
In our original paper investigating the figure of Fr\'aoch mac Fidach Foltruad from the Ulster Cycle~\cite{McCarthy2018}, we expressed a wish that it would ignite conversation between the disciplines and we believe that the fruits of that initial study were to the benefit of all concerned and will continue to flourish into the future. 
And so to finish, and in tribute to our colleague and friend, Professor Ralph Kenna, we will borrow a fitting quote from William S. Beck:

\begin{quotation}
	\centering{\emph{`Fields of learning are surrounded ultimately only by illusory boundaries-like the ``rooms'' in a hall of mirrors.\\
			It is when the illusion is penetrated that progress takes place. ... Likewise science cannot be regarded as a thing apart, to be studied, admired or ignored. It is a vital part of our culture, our culture is part of it, it permeates our thinking, and its continued separateness from what is fondly called ``the humanities'' is a preposterous practical joke on all thinking people.'}}\\
	\hspace{29em}{William S. Beck}
\end{quotation}

Ralph Kenna 1964--2023 --- `Ar dheis D\'e go raibh a anam'. 


\ukrainianpart

\title{На перехресті різних наук: початковий інструментарій для дослідників, які бажають досліджувати ранню ірландську літературу}
\author{М. Маккарті, Д. П. Керлі}
\address{
	Раткроганський історичний центр
}

\makeukrtitle

\begin{abstract}
	\tolerance=3000%
				Поштовхом для написання цієї статті стало довготривале співробітництво обох авторів із нашим другом та колегою, професором Ральфом Кенною.	Цей зв’язок спочатку виник завдяки його зацікавленості Раткроганом та нашою статтею ``Дослідження саги про Фраоха'', яку ми завершили твердженням, що, на нашу думку, вона ``представляє випадок, який започаткує дискусію між дисциплінами''. Це спонукало нас розглянути потенційну цінність укладання типового списку корисних та надійних джерел і ресурсів для дослідників. Іншими словами, певного початкового інструментарія або посібника для будь-якої особи --- представника альтернативної дисципліни або кваліфікації, у кого є (або виникне з часом) персональне чи професійне зацікавлення ранньою Ірландією та її літературою. Роблячи це, ми вирішили для простоти ілюстрації розглянути локації Раткроган та Круачан Еї (доісторичне королівське місце Коннахт на заході Ірландії та місце, де ми працюємо і де буваємо щодня) для тематичного дослідження, щоб продемонструвати початковий методологічний підхід не лише до певних доступних ресурсів та інформації, але й торкнутись деяких потенційних труднощів, що можуть виникнути під час наших студій.
	\keywords соцiальнi мережi, мiфологiя
	
\end{abstract}

\lastpage
\end{document}